\documentclass[twocolumn,preprintnumbers, amsmath, amssymb]{revtex4}

\def\be{\begin{equation}}
\def\ee{\end{equation}}
\def\beq{\begin{eqnarray}}
\def\eeq{\end{eqnarray}}
\def\n{\nonumber}
\def\bay{\begin{array}}
\def\eay{\end{array}}

\begin{document}
\preprint{CIRI/05-smw04}
\title{Heuristic approach to a {\em natural\/}
unification of the Quantum Theory and the General Theory of
Relativity}

\author{Sanjay M. Wagh}
\affiliation{Central India Research Institute, \\ Post Box 606,
Laxminagar, Nagpur 440 022, India\\
E-mail:cirinag\underline{\phantom{n}}ngp@sancharnet.in}

\date{September 12, 2004}
\begin{abstract}
In non-relativistic as well as in special relativistic quantum
theory, {\em mass\/} and {\em charge\/} are {\em pure numbers\/}
appearing in various (quantum) operators and admit {\em any
values}, {\it ie}, values for these quantities are to be
prescribed {\em by hand}. This is, in a theory of probability,
understandable since we need to {\em assume\/} some {\em
intrinsic\/} properties of the object we are calculating the
probability about. Then, if we {\em specify}, in some satisfactory
manner, mass and charge for a point of the space in a suitable
general-relativistic framework, the quantum theoretical framework
could, in principle, be {\em obtainable\/} within it. Heuristic
arguments are presented to show that a {\em natural unification\/}
of the quantum theory and the
general theory of relativity is achievable in this manner. \\

\centerline{To be submitted to: General Relativity \& Gravitation}
\end{abstract}
\maketitle

\newpage
Difficulties of cognition of physical phenomena arise only in
``constructing'' a theory from the results of experiments and
observations of the physical world, that is, when attempts are
made to establish a consistent cause and effect relation between
them. It might then seem necessary to demand that no concept enter
a theory which has not been experimentally established, at least
to the same degree of accuracy as the experiments to be explained
by that theory. This simple demand is, plainly speaking, quite
impossible \cite{heisenberg} to fulfil. It is therefore necessary
to introduce various {\em concepts\/} into a physical theory,
without justifying them rigorously, and then to allow the
experiments to determine the range of their applicability, to
decide at what points their revision is necessary.  Newton's
theory is, now, a classic example of this methodology of Physics.

Newton's ideas are geometric \cite{cartan} conceptions of {\em
specific mathematical structures or fields\/} (scalar, vector,
tensor functions) defined over the {\em metrically flat
(Euclidean) 3-continuum\/} and the laws of their spatio-temporal
transformations. The underlying metrically flat 3-continuum admits
the {\em same Euclidean metric structure\/} before and after the
(galilean) coordinate transformations. This ever-flat 3-continuum
is, in this sense, an {\em absolute space\/} in Newton's theory.
Furthermore, in Newton's theory, the physical laws for these
quantities are mathematical statements form-invariant under
galilean coordinate transformations.

Various ({\em newtonian\/}) mathematical methods, of Laplace,
Lagrange, Euler, Hamilton, Jacobi, Poisson and others
\cite{class-mech}, hold for these newtonian fields definable on
the metrically flat 3-continuum and are consistent with the fact
that the underlying flat 3-continuum admits the same metric
structure before and after the spatio-temporal transformations of
these fields. {\em This overall is, truly, the sense of any theory
being newtonian}.

The spatio-temporal (galilean) transformations under which the
newtonian laws are form-invariant are, as opposed to general, {\em
specific\/} transformations of coordinates. Newton's theory also
attaches {\em physical\/} meaning to the space coordinates and to
the time coordinate. In this theory, the space coordinate
describes the physical distance separating physical bodies and the
time coordinate describes the reading of a physical clock.

In addition to the above, the (newtonian) temporal coordinate has
{\em universally\/} the same value for all the spatial locations,
{\em ie}, all synchronized clocks at different spatial locations
show and maintain the same time. In other words, the newtonian
time coordinate is the {\em absolute\/} physical time.

The basic concept of this theoretical framework is, of course,
Newton's point-mass that moves along a one-dimensional curve of
the metrically ever-flat 3-continuum. In these above {\em
physical\/} associations of the newtonian theory, it is always
tacitly assumed that the interaction of a measuring instrument
(observer) and the object (a particle whose physical parameters
are being measured) is negligibly small or that the effects of
this interaction can be eliminated from the results of
observations to obtain, as accurately as desired, the values of
these parameters \cite{bohr1} \footnote{\label{foot1} This is
questioned in quantum theory. See, for example, Bohr
\cite{bohr1}.}.

An issue closely related to the above one is that of the
causality. Given initial data, Newton's theory predicts the values
of its variables of the point-mass exactly and, hence, assumes
strictly causal development of {\em its\/} physical world.

But, the question arises of zero rest-mass particles, if any, and
Newton's theory cannot describe motion of such particles for
obvious reasons:
\[\vec{a}=\frac{\vec{F}}{m}\] where $\vec{a}$ is the acceleration,
$\vec{F}$ is the force causing that acceleration of a newtonian
particle of an {\em inertial\/} rest-mass $m$. Clearly,
acceleration has no meaning for $m=0$ in the above equation of the
newtonian second law of motion. This above inability is an
indication of the limitations of the newtonian theoretical
framework.

Then, if a zero rest-mass particle were to exist in reality, and
nothing in Newton's theory prevents this, it is immediately clear
that we need to ``extend'' various newtonian conceptions.

Now, light displays phenomena such as umbra-penumbra, diffraction,
interference, polarization etc. But, Newton's corpuscular theory
needs {\em unnatural, non-universal}, inter-particle forces to
explain these phenomena. That light displays phenomena needing
{\em unnatural\/} explanations in Newton's theory could, with
hindsight, then be interpreted to mean that light needs to be
treated as a zero rest-mass particle.

Furthermore, as Lorentz had first realized very clearly, the
sources of the newtonian {\em forces\/} are the singularities of
the corresponding fields defined on the flat 3-continuum. Although
unsatisfactory, this nature of the newtonian framework causes no
problems of mathematical nature since this distinction is
maintainable within the formalism, {\em ie}, well-defined
mathematical procedures for handling this distinction are
possible.

Revisions of newtonian concepts were necessary by the beginning of
the 20th century. Firstly, efforts to reconcile some experimental
results with newtonian concepts failed and associated conceptions
led Einstein to Special Theory of Relativity \cite{ein-pop}.
Secondly, some other experiments, in particular, those related to
the wave-particle duality of radiation and matter, both, led to
non-relativistic quantum theory \cite{heisenberg}.

The methods of Non-Relativistic Quantum Field Theory
\cite{varadrajan} are also similar of nature to the above
newtonian methods in that these consider {\em quantum fields\/}
definable on the metrically flat 3-continuum. For these fields of
quantum character, we are of course required to modify the
newtonian mathematical methods. The Schr\"{o}dinger-Heisenberg
formalism achieves precisely this. Quantum considerations only
change the nature of the mathematical (field) structure definable
on the underlying metrically flat 3-continuum. That is,
differences in the newtonian and the quantum fields are
mathematically entirely describable as such. But, the metrically
flat 3-continuum is also, in the above sense, an absolute space in
these non-relativistic quantum considerations.

Now, importantly, the ``source properties'' of physical matter are
differently treated in the non-relativistic quantum field theory
than in Newton's theory. Mass and electric charge of a physical
body appear as pure numbers, {\em to be prescribed by hand\/} for
a point of the metrically flat 3-continuum, in Schr\"{o}dinger's
equation or, equivalently, in Heisenberg's operators. Quantum
theory then provides us the probability of the {\em location\/} of
the mass and the charge values in certain specific region of the
underlying metrically flat 3-continuum. Also, the numerical values
of the newtonian variables of the point, such as its linear
momentum, energy, angular momentum etc.\ are prescribed the
corresponding probabilities.

However, certain physical variables of the newtonian mass-point
acquire {\em discrete values\/} in the mathematical formalism of
the quantum theory. This discreteness of certain variables is the
genuine characteristic of the quantum theory and is a significant
departure from their continuous values in Newton's theory.

This quantum theory is fundamentally a theory that divides the
physical world into two parts, a part that is a system being
observed and a part that does the observation. Therefore, quantum
theory always refers to an {\em observer\/} who is {\em
external\/} to the system under observation. The results of the
observation, of course, depend in detail on just how this division
is made.

But, it must be recognized that classical concepts are not
completely expelled from the physical considerations in the
quantum theory. On the contrary, in Bohr's words, {\em it is
decisive to recognize that, however far the phenomena transcend
the scope of classical physical explanation, the account of all
evidence must be expressed in classical terms\/} \cite{bohr2}.
This applies in spite of the fact that classical (newtonian)
mechanics does not account for the observations of the
microphysical world. (Bohr offers ``complementarity of (classical)
concepts'' as an explanation for this.)

We then also note here that it is not possible to treat zero rest
mass particles in the non-relativistic quantum theory. As is well
known \cite{varadrajan}, Schr\"{o}dinger's equation or
Heisenberg's operators of this theory are meaningful only when
mass is non-vanishing: \[ \left[-\;\hbar^2\,
\frac{\nabla^2}{2m}+V\right] \Psi=\imath\hbar
\frac{\partial}{\partial t}\Psi
\] where $\Psi$ is Schr\"{o}dinger's $\Psi$-function
and other symbols have their usual meaning.

Clearly, the relevant operators are meaningful only for $m\neq 0$.
Essentially, it is the same limitation as that of Newton's theory.
Non-relativistic quantum field theory cannot therefore describe
phenomena displayed by light.

But, here, a physical body is described as a non-singular
point-particle, not as an extended object. That is, mass and
electric charge appearing herein are non-singularly defined only
for a point of the metrically always-flat 3-continuum.

Now, with Maxwell's electromagnetism \cite{jackson}, we realize
that the {\em particle\/} of electromagnetic radiation has zero
rest-mass - follows from the special relativistic mass-variation
with velocity. Then, in essence, the theory of {\em special
relativity enlarges the galilean group of transformations of the
metrically flat 3-continuum and time to the Lorentz group of
transformations of the metrically flat 3-continuum and time},
which is also treatable as a metrically flat Minkowski-continuum
\footnote{Nothing special about 4-dimensionality. It also existed
with Newton's theory. Differences between these two theories,
Newton's theory and the Theory of Special Relativity, arise from
only the kind of transformations that are being used by them.}.

Lorentz transformations keep Minkowski metric the same. Then,
special relativistic laws for electromagnetic fields (mathematical
structures on the metrically flat 4-continuum), Maxwell's
equations, are mathematical statements form-invariant under
Lorentz transformations.

With Maxwell's equations, we have the ``special relativistic laws
of motion'' for the sources and Maxwell's equations for the
(electromagnetic) fields. Then, as long as we treat the sources
and the fields separately, problems of mathematical nature do not
arise since well-defined mathematical procedures exist to handle
these concepts.

Standard mathematical methods then permit us again considerations
of classical fields on the metrically flat 4-continuum
\cite{class-mech}. The ``newtonian'' mathematical methods of
Hamilton-Jacobi-Poisson hold also for them, now in 4-dimensions,
and are consistent with the fact that the flat 4-continuum admits
the same metric structure before and after the Lorentz
transformations of these fields.

This above is, now, the sense of any theory being classical. The
underlying metrically flat 4-spacetime is then an {\em absolute
4-space\/} in the sense \footnote{Then, we note, in advance, that
Einstein's field equations of General Relativity are {\em not\/}
classical in the sense just described and Einstein did state this
fact in \cite{schlipp}.} described herein.

It may now be noted that, for zero rest mass particles, we can
ascribe vanishing rest mass to a point of the space (or the
spacetime manifold): \[  E^2=p^2+m^2  \] A point of the space then
has $m=0$ when $E=p$ and such a point necessarily moves with the
speed of light. Then, the Lorentz group allows form-invariant
Maxwell's equations describing massless electromagnetic radiation.

The Lorentz transformations under which special relativistic laws
are form-invariant are {\em specific\/} coordinate
transformations. Further, the concepts of {\em measuring rods\/}
and {\em clocks\/} are clearly subject to critical examination and
it then becomes clear that the ordinary newtonian these concepts
involve the tacit assumption that there exist, in principle,
signals that are propagated with an infinite speed. Then, as shown
by Einstein, the absolute character of time is now lost
completely: initially synchronized clocks at different spatial
locations do not keep the same time-value.

However, like with Newton's theory, coordinates have a direct
physical meaning in special theory of relativity: spatial
coordinates represent the length of a physical measuring rod and
temporal coordinate represents the time-duration measured by a
physical clock. Although it is the same association of physical
character, the Lorentz transformations constitute significant
departure from the newtonian concepts since time is no longer the
absolute time in special relativity.

But, in such (classical) special relativistic considerations, it
is always assumed that the effects of the interaction of a
measuring instrument (observer) and the object can be eliminated
from the results of observations to obtain, as accurately as
desired, the values of various parameters of the object
\footnote{\label{foot2} This is, of course, questioned in the
quantum theory. At a later stage, we reconsider penetrating
remarks by Bohr, Heisenberg and related issues.}. This is exactly
as it was in Newton's theory. Hence, (classical) special
relativity assumes strict causality.

Now, quantum fields require suitable equations that are
form-invariant under the Lorentz transformations to describe
quanta moving close to the speed of light in vacuum. These quantum
fields are, once again, mathematical structures definable on the
metrically ever-flat 4-dimensional (Minkowski) spacetime.

Methods of special relativistic quantum field theory \cite{dirac}
then handle such {\em quantum fields\/} on the metrically
ever-flat 4-continuum admitting a Minkowskian metric. For quantum
fields, we need to ``modify'' the classical newtonian mathematical
methods. These are appropriate generalizations of those of
Schr\"{o}dinger-Heisenberg methods. This, the
Dirac-Schwinger-Tomonaga formalism \cite{dirac}, achieves for the
metrically flat minkowskian 4-continuum that which the
Schr\"{o}dinger-Heisenberg formalism achieves for the newtonian
3-space and time. Then, the differences in the classical and the
quantum fields are mathematically entirely describable as such.
Non-relativistic results are recoverable when the velocities are
small compared to the speed of light.

However, the underlying Minkowski spacetime does not change under
the (Lorentz) transformations keeping the quantum equations
form-invariant and is also, in the earlier sense, an {\em absolute
4-space\/} here.

Once again, a physical body is represented in these special
relativistic quantum considerations by ascribing in {\em
non-singular\/} sense the mass and the charge as pure numbers to
points of the Minkowski 4-continuum in the corresponding
operators. The special relativistic quantum theory then provides
us the probability of the {\em spatial location and temporal
instant\/} of the mass and the charge values in a region of the
Minkowski 4-continuum, for all velocities limited by the speed of
light in vacuum. The numerical values of other variables of the
physical body are also prescribed the corresponding probabilities.

Other massless particles, {\it eg}, neutrinos, are also allowed in
the special relativistic quantum field theory due to the group
enlargement from that of the galilean group to the Lorentz group
of transformations.  It is this group enlargement that permits
form-invariant Dirac equation \cite{dirac}. This group enlargement
also permits us the theory of massive spin $\frac{1}{2}$ fermions:
as are electrons.

However, as noted before, there is nothing special about the
4-dimensional metrically ever-flat spacetime manifold and the same
physics can be described in a $3+1$ fashion without difficulties
of mathematical or physical nature.

But, there cannot be any possibility of explaining the origin of
``mass'' as well as of ``charge'' in this, quantum or not, special
relativistic theory, since these properties of a physical body are
the pure numbers to be prescribed {\em by hand\/} for the
corresponding (classical or quantum) mathematical description
\footnote{This will, in general, be true as long as there exists
some procedure for ``prescribing by hand'' these quantities in the
adopted mathematical formalism of any physical character. The
question then arises whether the formalism of the quantum theory
can provide a procedure of {\em not\/} prescribing these
quantities {\em by hand}.}. In other words, it is only after we
have {\em specified\/} the values of mass and charge for a source
particle that we can {\em obtain}, from the mathematical
formalisms of these theories, its further dynamics based on the
given (appropriate) initial data. Hence, the values of mass and
charge are not {\em obtainable\/} in these theories.

Clearly, therefore, some new developments are needed here to
account for the ``origin'' of such fundamental properties of
matter. It is then also clear that the newtonian and the special
relativistic frameworks, both, are not sufficiently general to
form the basis for the entire physics.

Now, Lorentz had clearly recognized \cite{subtle} (p. 155) the
notion of the {\em inertia of the electromagnetic field\/} that
follows from Maxwell's theory of electromagnetism: Maxwell's
theory shows that the electromagnetic field possesses inertia
which is not the same as that of its source particle.

Lorentz then had a clear conception that inertia (opposition of a
physical body to a change in its state of motion) could possess
{\em origin\/} in the field conception. Just as a person in a
moving crowd experiences opposition to a change in motion, a
particle (region of concentrated field) moving in a surrounding
field experiences opposition to a change in its state of motion.
This is Lorentz's conception of the field-origin of inertia.

Now, firstly, the distinction between the source and the field
must necessarily be obliterated in any formulation of this
conception. In other words, a field is the only basic concept and
a particle is a derived concept here. Secondly, the mathematical
formulation of this conception is also required to be {\em
intrinsically nonlinear}. Basically, in this conception, the field
cannot refer to some linear mathematical structure defined on the
non-dynamical metrically ever-flat continuum.

Clearly, solutions of linear equations, {\em eg}, those of
Maxwell's electromagnetism, obey superposition principle, and
required number of solutions can be superposed to obtain the
solution for any assumed field configuration. But, the sources
generating the assumed field configuration continue to be the
singularities of the field. Hence, there are no means here of
obliterating the distinction between sources and field since the
sources are the singularities of the field they generate.

Some non-linear field equations could conceivably possess
singularity-free solutions for the field. Solutions of such
(non-linear) field equations would also not obey the superposition
principle. Then, one could hope that these (non-linear) equations
for the field would permit some appropriate treatment of source
particles as singularity-free regions of concentrated field
energy.

An important question is now that of the appropriate (non-linear)
field equations of this overall nature, of obtaining these
equations without venturing into meaningless arbitrariness. In
fact, this question is not just that of the appropriate nonlinear
(partial) differential equations that could serve as the field
equations. Rather, this question is of some appropriate non-linear
mathematical formalism that need not even possess the character of
non-linear (partial) differential equations for the field as a
mathematical structure on the underlying continuum. (It is also
the issue \footnote{It is then definitely very interesting to note
here that Einstein, in the early 1940s, had shown interest
\cite{subtle} (p. 347) in this question.} of whether the most
fundamental formalism of physics could have a mathematical
structure other than that of the (partial) differential
equations.)

Historically, there did not exist with Lorentz any physical
guidelines for getting to these non-linear field equations. This,
very difficult and lengthy, path to these non-linear equations was
completely developed by Einstein alone.

Then, Einstein formulated \footnote{This 4-dimensional approach
appears ``natural'' because Lorentz transformations appear natural
in any comprehension of various physical implications of Maxwell's
equations, their invariance relative to the transformations of the
underlying metrically flat 4-continuum - the Minkowski continuum,
now. But, we must also note that the same can be treated,
equivalently, in a 3+1 fashion.} his ideas for a 4-continuum. The
pivotal point of his formulation of the relevant ideas is the
equivalence of {\em inertial\/} and {\em gravitational\/} mass of
a physical body, a fact known since Newton's times but which
remained only an assumption of Newton's theory.

On the basis of the {\em equivalence principle}, Einstein then
provided us the ``curved 4-geometry'' as a ``physically
realizable'' entity. His field equations provide then a way of
realizing these conceptions of a curved geometry physically.
Einstein's (makeshift) field equations of this General Theory of
Relativity are form-invariant under general (spacetime) coordinate
transformations - the principle of general covariance
\cite{std-texts}.

To arrive at his formulation of the general theory of relativity,
Einstein raised \cite{schlipp} (p. 69) the following two
questions: {\em Of which mathematical type are the variables
(functions of the coordinates) which permit the expression of the
physical properties of the space (``structure'')? Only after that:
Which equations are satisfied by those variables?}

He then proceeded to develop this theory in two stages, namely,
those dealing with
\begin{description} \item{(a)} pure gravitational field, and
\item{(b)} general field (in which quantities corresponding
somehow to the electromagnetic field occur, too).
\end{description}

The situation (a), the pure gravitational field, is characterized
by a symmetric (Riemannian) metric (tensor of rank two) for which
the Riemann curvature tensor does not vanish.

For the situation (b), Einstein \cite{schlipp} (p. 73) then set up
``preliminary equations'' to investigate the usefulness of the
basic ideas of general relativity: \[
R_{_{ij}}-\frac{1}{2}\,R\,g_{_{ij}}=-\,\kappa\, T_{_{ij}}\] where
$R_{_{ij}}$ denotes the Ricci tensor, $R$ denotes the Ricci
scalar, $\kappa$ denotes a proportionality constant and
$T_{_{ij}}$ denotes the energy-momentum tensor of matter. In these
equations, the energy-momentum tensor does not contain the energy
(or inertia) of the pure gravitational field.

In this connection, Einstein expressed \cite{schlipp} (p. 75) his
judgement and concerns about these preliminary equations as: {\em
The right side is a formal condensation of all things whose
comprehension in the sense of a field theory is still problematic.
Not for a moment, of course, did I doubt that this formulation was
merely a makeshift in order to give the general principle of
relativity a preliminary closed expression. For it was essentially
not anything {\em more\/} than a theory of the gravitational
field, which was somewhat artificially isolated from a total field
of as yet unknown structure}.

This general theory of relativity essentially frees Physics from
the association of physical meaning to coordinates and coordinate
differences, an assumption implicit in Newton's theory and in
special relativity. The formulation of Einstein's (makeshift)
field equations however attaches physical meaning to the invariant
distance of the curved spacetime geometry and considers it to be a
physically exactly measurable quantity.

Through these equations, geometric properties of the spacetime are
supposed to be determined by the physical matter. In turn, the
spacetime geometry is supposed to tell the physical matter how to
move. That is, the geodesics of the spacetime geometry are
supposed to provide the law of motion of the physical matter.

Now, we may imagine \cite{std-texts} a {\em small\/} perturbation
of the {\em background\/} spacetime geometry and obtain equations
governing these perturbations. We may also consider
\cite{std-texts} quantum fields on the {\em unchanging
background\/} spacetime geometry.

Then, such methods (of perturbative analysis and also of the
Quantum Field Theory in Curved Spacetime) are quite similar of
nature to methods adopted for either the flat 3-continuum or the
flat 4-continuum in that these consider ``mathematical fields''
definable on the fixed and metrically curved 4-continuum. Keeping
the nature of mathematical (field) structures the same, these
considerations only change the 4-space which is now a metrically
curved ``absolute'' 4-continuum.

But, as far as Lorentz's or Einstein's conceptions are concerned,
these above considerations of a quantum field theory in a curved
spacetime or perturbations of a curved spacetime geometry are,
evidently, {\em not\/} self-consistent since matter fields must
affect the background spacetime geometry. But, these are {\em
not\/} the real issues here.

Without going into further details of its formalism, we note that
General Theory of Relativity embodies another significant
departure from newtonian concepts since only the concept of {\em
coincidence\/} in spacetime is accepted uncritically in it
\cite{heisenberg}. At this point, it is then important to note
that this space-time coincidence is {\em usally\/} taken to be
that of a physical kind. That is, it is usually assumed that the
time coordinate is {\em measurable exactly\/} and so are the
spatial coordinates {\em measurable exactly\/} in a measuring
arrangement.

But, the interaction between observer and object is then assumed
to be negligibly small or that its effects can be eliminated from
the results of any measurement. This is then the situation with
the {\em standard formulation of general relativity}, that of
Einstein's (makeshift) field equations, even when it has finite
speeds for signal propagations. Clearly, these two are then
unrelated issues.

But, in microscopic phenomena, the interaction between observer
and object causes uncontrollable and large changes in the system
being observed. Then, every experiment performed to determine some
numerical quantity renders the knowledge of some others illusory,
since the uncontrollable changes caused to the observed system
alter the values of these other quantities.

Therefore, Einstein's field equations cannot be applied to
microscopic phenomena and, hence, in general, may only serve to
approximate the average behavior of physical phenomena. This much
is, perhaps, generally recognized.

However, what has mostly gone unnoticed is the fact that
Einstein's approach to his vacuum as well as (makeshift) field
equations with matter is beset with numerous internal
contradictions of serious physical nature \cite{smw-field}.

Firstly, Einstein's vacuum field equations are entirely
unsatisfactory \cite{smw-field} since these are field equations
for the pure gravitational field {\em without\/} even a
possibility of the equations of motion for the sources of that
field.

Certainly, matter cannot be any part of the theory of the vacuum
or the pure gravitational field. Then, there cannot be physical
objects in considerations of the pure gravitational field, except
as sources of such fields. Therefore, it obviously does not make
any sense whatsoever to say that the geodesics of such a
spacetime, geometry describing a pure gravitational field, provide
the law of motion for the material sources.

Now, a material particle is necessarily a spacetime singularity of
the pure gravitational field and, hence, mathematically, no
equations of motion for it are possible. Then, we have only
equations for the pure field but no equations of motion for the
sources creating those fields.

But, the vacuum field equations alone are {\em not enough\/} to
draw any conclusions of physical nature. Without the laws for the
motions of sources generating the (vacuum) fields, we have no
means of ascertaining or establishing the ``causes'' of motions of
sources. No conclusions of physical nature are therefore
permissible in this situation and, thus, the vacuum field
equations cannot lead us to physically verifiable predictions.

[Note that this above situation is markedly different from that
with special relativity. In special relativity, the background
geometry does not possess any {\em geometric singularity\/} at any
location, but only the (mathematical) fields defined on this
geometry can be singular. Then, similar to Newton's theory,
situations in special relativity lead us to physically testable
predictions.]

Secondly, Einstein's (makeshift) field equations with matter are
also not satisfactory \cite{smw-field} from the physical point of
view. Recall here that the energy-momentum tensor deals with
density and fluxes of particles. Then, unless a definition of what
constitutes a particle is, a-priori, available to us, we cannot
even construct the energy-momentum tensor for the physical matter.
Since a spacetime singularity cannot represent a physical
particle, the concept of a particle is {\em not\/} available to us
in Einstein's approach to makeshift field equations of general
relativity. Thus, the energy-momentum tensor is {\em not\/} any
well-defined concept and, with it, the field equations become
ill-posed.

What is of definite (historical too) importance is the fact that
Einstein had recognized many of these problems of physical nature
\cite{smw-field} with his (makeshift) approach. Of relevance here
are his penetrating remarks \cite{schlipp} (p. 675): {\em
Maxwell's theory of the electric field remained a torso, because
it was unable to set up laws for the behavior of electric density,
without which there can, of course, be no such thing as an
electromagnetic field. Analogously the general theory of
relativity furnished then a field theory of gravitation, but no
theory of the field-creating masses. (These remarks presuppose it
as self-evident that a field theory may not contain any
singularities, i.e., any positions or parts in space in which the
field-laws are not valid.)}

It is also of relevance to note here that Einstein's numerous
attempts at unified field theory did not impress others
\footnote{For example, Pauli \cite{subtle} (p. 347) wrote {\em
[Einstein's] never failing inventiveness as well as his tenacious
energy in the pursuit of [unification] guarantees us in recent
years, on the average, one theory per annum. ... It is
psychologically interesting that for some time the current theory
is usually considered by its author to be the ``definitive
solution.''}}. For example, Pauli demanded \cite{subtle} (p. 347)
to know what had become of the perihelion of Mercury, the bending
of light etc. Einstein was not overly bothered that there were no
good answer to these questions in his unified field theory. He
wrote ``Nearly all the colleagues react sourly to the theory
because it puts again in doubt the earlier general relativity.''
(Was he ready to {\em completely\/} abandon his {\em earlier\/}
general relativity?)

The cases mentioned by Pauli are, clearly, the ``pathological
cases'' in which related mathematical expressions for the
explanations of concerned phenomena ``agree'' with observations.
But, for the reasons mentioned earlier, the ``correct
explanations'' of the same cannot be those provided by the
makeshift field equations \cite{smw-field}. It is decisive to
recognize  this fact.

In Einstein's conceptions of dynamic curved geometry, the
continuum is no longer only an ``inert'' stage for the physical
fields, different mathematical structures on the continuum. {\em
The defining structure of the continuum must change as these
physical structures defined on it change}. In a sense, therefore,
the defining structure and the physical fields need to be {\em
inseparable\/} in any mathematical formulation of these ideas.

Then, to reemphasize the same point once again, we reiterate that,
as per Einstein's these conceptions of curved geometry and matter,
{\em the metric structure, the defining structure of the
continuum, must change as the physical fields on it change.
Therefore, what we need here is some mathematical formulation of
these ideas in which the physical fields and the metric structure
become essentially inseparable from each other}. (But, this above
is clearly not the situation with Einstein's makeshift field
equations with matter since matter-free equations are {\em
obtainable\/} from them \footnote{To quote Pais \cite{subtle} (p.
287): ``Einstein never said so explicitly, but it seems reasonable
to assume that he had in mind that the correct equations should
have no solutions at all in the absence of matter.'' It is but
clear now as to why this should be the case with the correct
equations. })

Now, as per Einstein's conceptions, {\em the space is to be
indistinguishable from physical bodies}. Thus, properties of
physical bodies are the properties of the space and vice versa. As
any physical body {\em changes}, these properties change, it then
also being a change in the structure of the space.

The question is now of suitable mathematical description of these
(Einstein's and Descartes's) conceptions \cite{ein-pop}. On very
general grounds, such a mathematical description can be expected
to possess the following characteristics.

Clearly, in this description, motion of a physical body will be a
change in the structure of the space. Cartan's volume-form should
then be well-defined at every spatial location. A point of space
could also be prescribed, in some suitable non-singular sense, the
inertia, electric charge etc. Then, such a point of the space is
describable as a point particle in the newtonian sense. We may
also look at any {\em extended\/} physical body in the newtonian,
non-singular, sense of a point particle possessing various
properties of a physical body. Thus, physical bodies should {\em
everywhere in space\/} be describable as singularity-free.

The issue is of incorporating time in this framework and, the
temporal evolution of ``points of space'' is a mathematically
well-definable concept - as a dynamical system.

Then, concepts under consideration could be realizable in some
mathematically well-defined formalism that deals with dynamical
systems defined on continuum as the underlying set.

Such a description then also follows the principle of general
covariance: the laws regarding physical objects in it are based on
the arbitrary transformations of coordinates of the underlying
space and also on time as an essentially arbitrary parameter of
the dynamical system.

The question now is of suitable mathematical structure on the
space that allows us the association of physical properties of
material objects to the points of the space. Furthermore, the
question is also of defining in a natural manner the boundary of
any physical object.

A physical object has associated with it various (fundamental)
physical properties, {\em eg}, (rest) energy. Then, the adjacent
objects clearly separate by boundaries at which the spatial
derivative(s) of that property under consideration, ({\em eg},
rest energy), change(s) the sign.

But, any physical object is some {\em region of space}. Therefore,
some suitable structure on the space must, as per these
conceptions, then possess a similar property of its derivative
changing its sign at a boundary of a physical object.

It then also follows that a physical property can, essentially, be
specified {\em independently\/} for each spatial direction since
these directions are to be treated as {\em independent\/} of each
other. This is an immediate consequence of the fact that, as per
Einstein's conceptions, properties of the space are also the
properties of physical objects.

Now, the space (continuum) is characterizable by ``distance''
separating its points. Suitable ``distance'' function can then be
expected \footnote{This is a ``necessary'' consequence of these
conceptions. We can, of course, choose the defining property (of a
physical object) itself as a ``distance'' function.} to possess
the property of its derivative(s) changing sign across boundaries
separating regions of space corresponding to separated objects.
This suitable distance function then, mathematically, becomes a
pseudo-metric function on the space, remaining a metric function
within a region.

An obvious question is then of suitable such pseudo-metric
function on the underlying space, evidently, a continuum.

At the present juncture, we seek an answer to this question in the
properties of physical matter. Then, we note that physical matter
can be assembled (and reassembled) to produce (another form of)
physical matter. Since there is to be ``no space without physical
matter'' in the present situation, it then follows that the
underlying space must be mass-scale independent and, hence,
spatial-scale independent as well. This, in general, leads us to a
space admitting three linearly independent (spatial) homothetic
Killing vectors \cite{smw-issues}.

Consider therefore a three-dimensional pseudo-Riemannian manifold,
denoted as ${\rm \bf B}$, admitting a pseudo-metric
\cite{smw-issues}: \setcounter{equation}{0} \beq \label{3-metric}
d\ell^2= {P'}^2Q^2R^2\, dx^2 &+&P^2\bar{Q}^2 R^2\, dy^2 \n \\
&+&\;P^2Q^2\tilde{R}^2 \,dz^2 \eeq where we have $P\equiv P(x)$,
$Q\equiv Q(y)$, $R\equiv R(z)$ and $P'=dP/dx$, $\bar{Q}=dQ/dy$,
$\tilde{R}=dR/dz$. The vanishing of any of these spatial functions
is a {\em curvature singularity}, and constancy (over a range) is
a {\em degeneracy\/} of (\ref{3-metric}).

A choice of functions, say, $P_o$, $Q_o$, $R_o$ is a specific
distribution of ``physical properties'' in the space of
(\ref{3-metric}). As some ``region'' of physical properties
``moves'' in the space, we have the original set of functions
changing to the ``new'' set of corresponding functions, say,
$P_1$, $Q_1$, $R_1$.

Clearly, we are considering the isometries of (\ref{3-metric})
while considering ``motion'' of this kind. Then, we will remain
within the group of the isometries of (\ref{3-metric}) by
restricting to the triplets of {\em nowhere-vanishing\/} functions
$P$, $Q$, $R$. We also do not consider any degenerate situations
for (\ref{3-metric}).

If we denote by $\ell$ the pseudo-metric function corresponding to
(\ref{3-metric}), then $({\rm \bf B},\ell)$ is an uncountable,
separable, complete pseudo-metric space. If we denote by $d$, a
metric function canonically \cite{kdjoshi} obtainable from the
pseudo-metric (\ref{3-metric}), then the space $({\rm\bf B}, d)$
is an uncountable, separable, complete metric space. If $\Gamma$
denotes the metric topology induced by $d$ on ${\rm \bf B}$, then
$({\rm\bf B}, \Gamma)$ is a Polish topological space. Further, we
also obtain a Standard Borel Space $({\rm\bf B},\mathcal{B})$
where $\mathcal{B}$ denotes the Borel $\sigma$-algebra of the
subsets of ${\rm\bf B}$, the smallest one containing all the open
subsets of $({\rm\bf B},\Gamma)$ \cite{trim6}.

Since $({\rm\bf B},\mathcal{B})$ is a standard Borel space, any
measurable, one-one map of ${\rm\bf B}$ onto itself is a Borel
automorphism. Therefore, the Borel automorphisms of $({\rm\bf
B},\mathcal{B})$, forming a group, are natural for us to consider
here.

But, the pseudo-metric (\ref{3-metric}) is a metric function on
certain ``open'' sets, to be called the P-sets, of its Polish
topology $\Gamma$. A P-set of $({\rm\bf B},d)$ is therefore never
a singleton subset, $\{ \{x\}:x\in {\rm\bf B}\}$, of the space
${\rm\bf B}$. Note also that every open set of $({\rm\bf B},
\Gamma)$ is {\em not\/} a P-set of $({\rm\bf B}, d)$.

Any two P-sets, $P_i$ and $P_j$, $i,j\;\in\;{\rm\bf N}$,
$i\,\neq\,j$, are, consequently, {\em pairwise disjoint subsets\/}
of ${\rm\bf B}$. Also, each P-set is, in own right, an
uncountable, complete, separable, metric space.

Evidently, {\em a P-set is the mathematically simplest form of
``localized'' physical properties in the space ${\rm\bf B}$\/} and
we call it a {\em physical particle}. This suggests that suitable
mathematical properties of a P-set, as well as those of a
collection of P-sets, can be the properties of physical matter.

Now, {\em (Lebesgue) measures\/} on {\em measurable\/} subsets of
a standard Borel space are {\em natural\/} for us to consider
here. Also, {\em signed\/} measures are definable on measurable
sets. Signed measures then provide us the notion of the
``polarity'' of certain properties. {\em Measures\/} could then
represent {\em physical propoerties\/} in the present formalism.

Now, P-sets, also open in $\Gamma$, are {\em measurable\/} subsets
of ${\rm\bf B}$. Thus, we associate with every attribute of a {\em
physical body}, a suitable class of (Lebesgue) measures on such
P-sets. Hence, a P-set is a {\em physical particle}, always an
{\em extended body}, since a P-set cannot be a singleton subset of
${\rm\bf B}$.

Clearly, various physical properties (measures) {\em change\/}
only when the region of space (P-set) changes. Thus, a region of
space (P-set) and physical properties (the measures on P-sets)
have been amalgamated into one thing here.

Moreover, a given measure can be integrated over the underlying
P-set in question. The integration procedure is always a
well-defined one for obvious mathematical reasons. The value of
such an integral provides then an ``averaged quantity
characteristic of a P-set'' under question. It is then evident
that this ``average'' is a property of the entire P-set under
consideration and, hence, of {\em every point\/} of that P-set.

A point of the P-set is then thinkable as having these averaged
properties of the P-set and, in this precise non-singular sense,
is thinkable as a (newtonian) {\em point-particle\/} possessing
those averaged properties. In this non-singular sense, points of
the space ${\rm\bf B}$ become point particles.

In essence, we have,  in a non-singular manner, then ``recovered''
the (newtonian) notion of a point particle from that of our notion
of a field - the underlying continuum ${\rm\bf B}$.

Further, the ``location'' of this point-particle will be {\em
indeterminate\/} over the {\em size\/} of that P-set because the
averaged property is also the property of every point of the set
under consideration. The individuality of a point particle is then
that of the corresponding P-set.

Now, we call as {\em an object\/} a collection of P-sets. But, a
P-set is a particle. Therefore, an {\em object\/} is a {\em
collection\/} of particles. Measures can also be integrated over
such objects. Such integrated measures are then the property of
every point of that object under consideration and a point of
${\rm\bf B}$ in the object is then also thinkable as a (newtonian)
point particle with these physical properties \footnote{That we
are able to perform these operations is to be attributed to the
mass-scale or length-scale invariance.}. Location of such a point
particle is then indeterminate over the size of that object.

Then, the points of the underlying space ${\rm\bf B}$ can also be
attributed the physical properties averaged over the size of an
object. Hence, we can also represent an object under consideration
as a (newtonian) point particle.

Therefore, we have the required characteristics of Descartes's and
Einstein's conceptions incorporated in the present formalism.
Clearly, we have then the non-singular notion of a point particle
as well as that of replacing any extended physical body by such a
non-singular point particle. Furthermore, physical bodies are also
represented as non-singular regions of the space ${\rm\bf B}$.
Then, the union of the space and the physical objects is clearly
perceptible here.

As any Borel automorphism of the underlying space ${\rm\bf B}$
changes a P-set/object, the integrated properties may also change
and, hence, the (initial) particle(s) may change into other
particle(s), if integrated measures change.

Now, the {\em Hausdorff metric\/} \cite{kdjoshi} provides the
distance separating P-sets and also the distance separating
objects. This distance bewteen sets will, henceforth, be called
the {\em physical distance\/} between P-sets or objects (as
extended physical bodies) because ``measurement in the physical
sense'' can be expected to yield only this quantity as distance
separating physical objects.

Measure-preserving Borel automorphisms of the space ${\rm\bf B}$
then ``transform'' a P-set maintaining its characteristic classes
of (Lebesgue) measures, that is, its physical properties.

Non-measure-preserving Borel automorphisms change the
characteristic classes of Lebesgue measures (physical properties)
of a P-set while ``transforming'' it. Evidently, such
considerations also apply to objects.

Then, a {\em periodic\/} or {\em periodic component of\/} Borel
automorphism \cite{trim6} will lead to an {\em oscillatory
motion\/} of a P-set or an object while preserving or not
preserving its measures.

Therefore, an object undergoing periodic motion is a physically
realizable clock in the present framework. Such an object
undergoing oscillatory motion then ``displays'' the time-parameter
of the corresponding (periodic) Borel automorphism since the
period of the motion of such an object is precisely the period of
the corresponding Borel automorphism. ({\em It is however not the
``observed'' time since we `need' here the means to `observe' the
periodic motion of the object}.)

Then, within the present formalism, a {\em measuring clock\/} is
therefore any P-set or an object undergoing {\em periodic motion}.
A P-set or an object can also be used as a {\em measuring rod}.

Therefore, in the present theoretical framework, measuring
apparatuses, measuring rods and measuring clocks, are on par with
every other thing that the formalism intends to treat.

A Borel automorphism of $({\rm\bf B},\mathcal{B})$ may change the
physical distance resulting into ``relative motion'' of objects.
We also note here that the sets invariant under the specific Borel
automorphism are characteristic of that automorphism. Hence, such
sets will then have their distance ``fixed'' under that Borel
automorphism and will be stationary relative to each other.

Now, in a precise sense, it follows that the position of the
point-particle (of integrated characteristics of its P-set) is
``determinable'' more and more accurately as the size of that
P-set gets smaller and smaller. But, complete localization of a
point particle is not permissible here since a P-set or an object
is never a singleton subset of ${\rm\bf B}$. The location of the
point particle is then always ``indeterminate'' to the extent of
the size of its P-set. This is an intrinsic indeterminacy that
cannot be overcome in any manner.

Hence, a joint manifestation of Borel automorphisms of the space
$({\rm\bf B}, \mathcal{B})$ and the association, as a point
particle, of integrated measures definable on a P-set or an object
with the points of ${\rm\bf B}$ is a candidate reason behind
Heisenberg's indeterminacy relations since indeterminacy of
location of a point-particle is intrinsic here.

In this context, we therefore note that any experimental
arrangement to determine a physical property of a P-set or an
object is based on some specific ``arrangement'' of P-sets and
involves corresponding Borel automorphisms of ${\rm\bf B}$
affecting those P-sets or objects.

For example, Heisenberg's microscope attempting the determination
of the location of an electron involves the collision of a photon
with an electron. It therefore has an associated Borel
automorphism producing the motion of a specific P-set, a photon.
Although we have not specified the sense in which a P-set can be a
photon, it is clear that the Borel automorphism causing its motion
will also affect an electron as a P-set.

Thus, a P-set ``transforms'' as a result of our efforts to
``determine any of its characteristic measures'' since these
``efforts or experimental arrangements'' are also Borel
automorphisms, not necessarily the members of the class of Borel
automorphisms keeping  invariant that P-set (as well as the class
of its characteristic measures).

Hence, a Borel automorphism (experimental arrangement)
``determining'' a characteristic measure of a P-set changes, in
effect, the very quantity that it is trying to determine. This
peculiarity then leads to Heisenberg's (corresponding)
indeterminacy relation.

As another example, consider a physical clock. In the present
context, it is an object undergoing periodic motion, with the
period of its motion being {\em exactly\/} that of the
corresponding periodic Borel automorphism. However, to be able to
determine this time period of the physical clock, we need the
``interaction'' of another agency, say, a photon, with the clock,
and that interaction is another Borel automorphism. This
``interaction'' of observing agency causes the ``change'' in the
period of the clock and, hence, leads to the corresponding
indeterminacy relation.

Once again, we have not specified the sense in which a P-set can
be a photon here.  However, it follows that this ``sense'' is
precisely that of measures definable on a P-set of the underlying
continuum space ${\rm \bf B}$. Here, we therefore need to ``define
clearly'' the classes of measures corresponding to a photon and an
electron and, we have obviously not done that here.

However, irrespective of this obvious question of only
mathematical nature, it is clearly possible to intuitively explain
\cite{smw-indeterminacy} the origin of Heisenberg's indeterminacy
relations. The formalism of dynamical systems on the Borel space
${\rm\bf B}$ then provides us therefore an ``origin'' of
Heisenberg's indeterminacy relations. This is in complete contrast
to their probabilistic origin as advocated by the standard
formalism of the quantum theory.

Notice now that, in the present considerations, we began with none
of the fundamental considerations of the concept of a quantum.
But, one of the basic characteristics of the conception of a
quantum, Heisenberg's indeterminacy relation, emerged out of the
present formalism.

Furthermore, in the present framework, we have also done away with
the ``singular nature'' of the particles and, hence, also with the
unsatisfactory dualism of the field (space) and the source
particle. We also have, simultaneously, well-defined laws of
motion (Borel automorphisms) for the field (space) and also for
the well-defined conception of a point particle (of integrated
measure characteristics of a P-set or an object). Then, the
present formalism is a {\em complete\/} field theory.

However, none of the two notions of location and momentum is any
deficient for a description of the facts since Heisenberg's
indeterminacy relations are also ``explainable'' within the
present formalism. This explanation crucially hinges on the fact
that the points of the space ${\rm\bf B}$, as singleton subsets of
the space ${\rm\bf B}$, are never the P-sets. It is only in the
sense of associating the measures integrated over a P-set that the
points of the space ${\rm\bf B}$ are point particles.

[At this point, we then also note that the Borel automorphisms of
${\rm\bf B}$ need not be differentiable or, for that matter, even
continuous. Therefore, the present considerations also use, for
the most fundamental formalism of physics, a mathematical
structure different than that of the partial differential
equations. However, the question of the physical significance of
non-differentiable and non-continuous Borel automorphisms of
${\rm\bf B}$ is a subject of independent detailed study.]

Now, any measuring arrangement is conceivable here only as a Borel
automorphism of the underlying continuum ${\rm\bf B}$. It is
therefore clear that the measurability of any characteristic of a
point particle as defined in the present framework is dependent on
the ``Borel automorphism'' to be used. But, that Borel
automorphism changes the very P-set or object of measurement.

The ``determined or observed'' location of a particle is therefore
a {\em different\/} conception here. It clearly depends on the
Borel automorphism to be used for the measurement. The ``observed
velocity or momentum'' of a particle is then a conception
dependent on the notion of the physical distance changing under
the action of a Borel automorphism of ${\rm\bf B}$. Clearly, the
coordination of the underlying continuum ${\rm\bf B}$ has nothing
whatsoever to do with the measurability here.

This brings us to some most fundamental issues of the physical
world, those of the physical description, measurement and
causality.

Recall what Bohr \cite{bohr1} had put it so succinctly: {\em The
quantum theory is characterized by the acknowledgement of a
fundamental limitation in the classical physical ideas when
applied to atomic phenomena. ... the so-called quantum postulate,
which attributes to any atomic process an essential discontinuity,
or rather individuality, completely foreign to the classical
theories. ...}

Bohr \cite{bohr1} continues: {\em This postulate implies a
renunciation as regards the causal spacetime coordination of
atomic processes. Indeed, our usual description of physical
phenomena is based entirely on the idea that the phenomena
concerned may be observed without disturbing them appreciably.
This appears, for example, clearly in the theory of relativity,
which has been so fruitful for the elucidation of the classical
theories. As emphasized by Einstein, every observation or
measurement ultimately rests on the coincidence of two independent
events at the same spacetime point. Just these coincidences will
not be affected by any differences which the spacetime
coordination of different observers may exhibit. Now, the quantum
postulate implies that any observation of atomic phenomena will
involve an interaction with the agency of observation not to be
neglected. Accordingly, an independent reality in the ordinary
physical sense can neither be ascribed to the phenomena nor to the
agencies of observation. ...}

Indeed, the quantum theory has taught us that the interaction of
an observing mechanism with the object being observed causes
``uncontrollable'' changes to that object of observation. This is
an entirely {\em new element\/} that was not present either in
Newton's theory or in special relativity. This {\em new element\/}
definitely signifies a fundamental limitation of the related
ideas.

To illustrate this above issue, Heisenberg \cite{heisenberg} thus
wrote that: {\em In fact, our ordinary description of nature, and
the idea of exact laws, rests on the assumption that it is
possible to observe the phenomena without appreciably influencing
them. To coordinate a definite cause to a definite effect has
sense only when both can be observed without introducing a foreign
element disturbing their interrelation. The law of causality,
because of its very nature, can only be defined for isolated
systems, and in atomic physics even approximately isolated systems
cannot be observed.}

In his exposition of Bohr's concept of complementarity, Heisenberg
\cite{heisenberg} then, rightfully, pointed out that: {\em Second
among the requirements traditionally imposed on a physical theory
is that it must explain all phenomena as relations between objects
existing in space and time. This requirement has suffered gradual
relaxation in the course of the development of physics.}

In this connection, Bohr \cite{bohr1} wrote that: {\em This
situation has far reaching consequences. On one hand, the
definition of the state of a physical system, as ordinarily
understood, claims the elimination of all external disturbances.
But, in that case, according to the quantum postulate, any
observation will be impossible, and, above all, the concepts of
space and time lose their immediate sense. On the other hand, if
in order to make observation possible we permit certain
interactions with suitable agencies of measurement, not belonging
to the system, an unambiguous definition of the state of the
system is naturally no longer possible, and there can be no
question of causality in the ordinary sense of the word. ...}

He added \cite{bohr1} further that: {\em Just as the relativity
theory has taught us that the convenience of distinguishing
sharply between space and time rests solely on the smallness of
the velocities ordinarily met with compared to the velocity of
light, we learn from the quantum theory that the appropriateness
of our usual causal spacetime description depends entirely upon
the small value of the quantum of action as compared to the
actions involved in ordinary sense perceptions.}

Heisenberg \cite{heisenberg} too had, similarly, stated that: {\it
Although the theory of relativity makes the greatest of demands on
the ability for abstract thought, still it fulfills the
traditional requirements of science in so far as it permits a
division of the world into subject and object (observer and
observed) and hence a clear formulation of the law of causality.
This is the very point at which the difficulties of the quantum
theory begin.}

Furthermore, Bohr \cite{bohr1} also mentioned: {\em Strictly
speaking, the idea of observation belongs to the causal spacetime
way of description. ... According to the quantum theory, just the
impossibility of neglecting the interaction with the agency of
measurement means that every observation introduces a new
uncontrollable element.}

Surely, there is absolutely no way of going back to the newtonian
strict causality {\em simultaneously\/} with its associations of
physical properties of objects and the underlying space.

Now, these references to ``the relativity theory'' are clearly
directed at special relativity and emphasize that the Lorentz
transformations involve space and time coordinates on equal
footing in the 4-dimensional Minkowski manifold.

But, these considerations then ignore the fact that there is,
physically speaking, really nothing sacrosanct about the
4-dimensionality in special relativity. Then, we also note at this
place that no adequate recognition of Einstein's and Descartes's
ideas (about physical matter and geometry) is evident in these
above remarks as well as in related discussions by Bohr and by
Heisenberg, both. These ideas have definitely changed the
newtonian concept of ``all phenomena as relations between objects
existing in space and time'' since their conceptual framework is
based on the inseparable unison of physical objects and the space
- physical objects are the space and vice versa.

For this last reason, the issues involved herein may also {\em
not\/} be so simple and straightforward as appears from the above
remarks \cite{heisenberg, bohr1}. This situation forces us to
reconsider some contents, while still maintaining the validity of
other contents, of these remarks \cite{heisenberg, bohr1}.

Firstly, it is vital to recognize that the aforementioned {\em new
element\/} has a direct bearing only on the ``observability'' of
involved quantities but the ensuing limitation does {\em not\/}
apply to the {\em association\/} of the properties  of a physical
body with the properties of the space. Observability and these
associations are independent issues.

To illustrate this point further, let us consider that an observer
chooses a certain spatial location as the origin of the coordinate
system and intends to attach {\em spatial labels\/} to other
points of the space with respect to it. To achieve this, the
observer needs to use a {\em measuring rod}.

The observer intends to also attach suitable {\em temporal
labels\/} to each point of the space. For this purpose, the
observer then needs to place at each location in the space
physical bodies executing periodic motion as {\em clocks}.

But, by the quantum postulate, these concepts involve {\em
intrinsic\/} indeterminacies. Therefore, neither the origin of the
coordinate system nor the coordinate labels can be determined in a
{\em physical measurement\/} any more accurately than permitted by
Heisenberg's relevant indeterminacy relation. It is therefore that
the coordination used by an observer is not the same as the
coordination of the underlying continuum.

Then, in our considerations, the Hausdorff metric, and not the
pseudo-metric (\ref{3-metric}), is clearly the only which is
relevant to the coordination used by an {\em observer}. That is to
say, the coordination of the space ${\rm\bf B}$ has no direct
bearing on the coordination used by an observer.

Furthermore, the association of physical properties, measures on
the P-sets or objects, with the underlying space, the space
${\rm\bf B}$, has nothing whatsoever to do with the
``observability'' of these properties by an observer.
Observability rests, clearly, on the (group of) Borel
automorphisms of the underlying continuum ${\rm\bf B}$.
Observability and the association of physical properties with the
underlying space are therefore entirely unrelated issues for our
considerations here. It is clearly decisive to recognize this
independence of these issues.

Now, physical measurement involves the coincidence of two
``events'' at the {\em same space\/} location as far as the
coordination of the space ${\rm\bf B}$ is concerned and at the
same ``time'' as far as the ``labelling parameter'' of the
corresponding Borel automorphism of the space ${\rm\bf B}$ is
concerned. But, no observer can bypass the natural limits
specified by the indeterminacy relations.

Hence, Einstein's point that every measurement rests on the
coincidence of two independent events at the same spacetime point
should really be viewed as a statement about the coordination of
the underlying continuum ${\rm\bf B}$. Then, it also follows that
these coincidences will not be affected by the differences in the
spacetime coordination used by different observers.

Then, Einstein's arguments \cite{bohr1} about the spacetime
coincidence of two events apply in that two independent events, as
far as the coordination of the underlying continuum is concerned,
occur at the same space point and at the same time. But, Bohr
\cite{bohr1} is equally well justified in pointing out that the
quantum postulate renders ``measurability'' of the values of
(spacetime) coordinates {\em indeterminate\/} within the natural
limits specified by Heisenberg's indeterminacy relations. This
dichotomy is evident in the formalism here.

Presently, it is equally vital to also recognize that {\bf\it the
basic ideas of general relativity, of the inseparable unison of
physical matter and curved geometry, do {\em not\/} imply any
assertion that its coordinates, coordinate differences and even
the invariant geometric distance are \/{\em exactly measurable\/}
physical quantities}. Then, the concepts of underlying curved
geometry are {\em independent\/} of those of physical
measurements. The physical meaning attached to the geometric
invariant distance is then, decisively, also lost.

Importantly, the present formalism of dynamical systems on space
${\rm\bf B}$ follows the ``strict determinism'' in the obvious
sense that the space coordinates map {\em exactly\/} under the
action of a Borel automorphism of the space ${\rm\bf B}$.

However, it is equally important to realize that this is {\em
not\/} the same ``causality'' as that of the newtonian physical
formulation since the present formulation must necessarily satisfy
Heisenberg's indeterminacy relations.

In this connection, we note that the strict newtonian causality
implies that given precise position and velocity of a particle at
a given moment and all the forces acting on it, we can predict the
precise position and velocity of that particle at any later moment
using appropriate (classical) laws - Newton's laws. Then, this
newtonian causality is gone for good with the realization that the
quantum postulate \cite{bohr1} must hold always.

In our present formalism, the newtonian causality would have
demanded that the position and the velocity of a point-particle
(definable in the present formalism as a point of the P-set or an
object with associated integrated measures) be exactly
determinable. This is, evidently, not the situation in the present
formalism of dynamical systems on the underlying continuum
${\rm\bf B}$.

But, {\em reality independent of any act of observation}, is then
ascribable to the phenomena as well as to the agencies of
observation in this formalism. Since physical objects are the
regions of the space and vice versa, the ``existence'' of the
points of the space is the ``existence'' of physical objects. This
``existence'' is, obviously, independent of any act of
observation. {\em Objective reality\/} of physical phenomena is
then explainable similarly. That this is doable in a
mathematically and physically consistent manner is now evident.

Clearly, the Moon of the Mother Earth then ``exists'' whether we
choose to look at it or not! But, evidently now, by choosing to
look at it, we do also change the gross or average physical
characteristics of the Moon as well!

Furthermore, to coordinate a definite cause to a definite effect
has appropriate sense in the present formalism. The Borel
automorphism of the space ${\rm\bf B}$ is the {\em cause\/} behind
a certain effect on the P-set or an object. It is therefore not
that the idea of {\em exact laws\/} rests entirely on the
assumption that it is possible to observe the phenomena without
appreciably influencing them.

Then, the present formalism of dynamical systems on the underlying
continuum ${\rm\bf B}$ allows us a clear division of the world
into act of observation and observed. It is this subject-object
division that is behind the {\em law of causality\/} of the
present formalism. But, resolutions of the difficulties of the
quantum are already incorporated into the present formalism in
that the associations of physical properties with the underlying
space ${\rm\bf B}$ embody the indeterminacy relations.

{\em The present formalism, of the dynamical systems of the
underlying continuum ${\rm\bf B}$, is then, already, a unification
of the quantum theory and the general theory of relativity.}

[By the ``General Theory of Relativity,'' we of course do not mean
here the standard formulation in terms of the Einstein makeshift
field equations. We, rather, mean here the underlying ideas of the
inseparable unison of physical objects and the space which
necessarily require the {\em generality\/} of coordinate
transformations.

Similarly, by the ``quantum theory,'' we do not mean here the
standard formulation in terms of the $\Psi$-function but only the
underlying ideas implied by Bohr's quantum postulate.

It is a separate and important issue as to how one would recover
these standard formulations from the present framework of
dynamical systems of the space ${\rm\bf B}$.]

In what follows, we now provide further heuristic reasons as to
why this unification can be considered ``natural'' in a definite
sense.

Recall that the standard formalism of the quantum theory based on
Schr\"{o}dinger's $\Psi$-function provides us, essentially, the
means of calculating the probability of a physical event involving
physical object(s). It presupposes that we have specified, say,
the lagrangian or, equivalently, certain physical characteristics
of the problem under consideration. Evidently, this is necessary
to determine the $\Psi$-function using which we then make
(probabilistic) predictions regarding that physical phenomenon
under consideration.

At this stage, we then note the following fundamental limitation
of any theory that uses probabilistic considerations. (This
limitation is clearly recognizable for statistical mechanics in
relation to the newtonian theory.)

The pivotal point here is that the method of calculating the
probability, say, of the outcome of the toss of a coin, {\em per
se}, cannot provide us the physical description or
characterization of what we mean by that coin, whose certain {\em
intrinsic\/} properties must be specified before hand. That is to
say, the method of obtaining the probability of the outcome of its
toss is irrelevant to certain intrinsic properties of the coin.

Therefore, methods of the standard quantum theory, these leading
us to the probability of the outcome of a physical experiment
about a chosen physical object, cannot provide us the means of
``specifying'' certain intrinsic properties of that physical body.
This fact, precisely, appears to be the reason as to why we had to
{\em specify by hand\/} the values of the mass and the charge in
various operators of the non-relativistic as well as relativistic
versions of the quantum theory.

The ``origins'' of ``intrinsic'' properties of physical bodies
cannot then be explainable on the basis of the $\Psi$-function or
by employing the standard methods of the quantum theory.

It is then equally decisive to recognize also that this {\em
intrinsic\/} limitation of probabilistic considerations cannot, in
any manner, be circumvented by {\em redefining\/} probability
\footnote{As a historical interest, we then note here that, during
the entire period of the development of the standard formalism of
the quantum theory, Einstein had showed \cite{subtle} no interest
in attempts at redefining probability.}.

Essentially therefore, the standard formulation of the quantum
theory {\em presupposes\/} then that we have {\em specified\/}
certain intrinsic properties of physical object(s) under
consideration. Hence, the origin of such properties is to be
sought elsewhere and not within the formalism of the quantum
theory. [This is then also why the formalism of quantum theory as
well its inherently probabilistic considerations cannot provide
the {\em universal\/} basis for physics, in general.]

If not within the standard formalism of the quantum theory then,
how else can we ``explain'' the ``origin'' of the intrinsic
properties of material bodies such as its inertia?

In this connection, it is but clear that, in any such explanation,
we will have to employ the concept of the space which has been the
basis of physical theories since Newton's times. This is necessary
not just for the reasons of being able to connect with the ideas
of the past but also for the fact that no other avenues appear to
be available. Then, Descartes's and Einstein's ideas of the
inseparable unison of physical bodies and the properties of the
space become open for exploration.

Of course, there cannot be ``source particles'' as separate or
fundamentally independent physical entities in a complete field
theory. Particles can only be represented as special regions of
field energy in the space and, as per Descartes's and Einstein's
ideas, the space itself is the field. The properties of the space
are then the properties of material bodies and vice versa. These
conceptions, together with certain physically motivated
observations related to the mass-scale or the length-scale
independence of the space, then lead us uniquely to the
pseudo-metric space ${\rm\bf B}$.

We are then led to the approach of {\em measures\/} definable on
the pseudo-metric 3-continuum ${\rm\bf B}$ to represent points of
the continuum as particles with properties as the integrated
measures defined on the corresponding P-sets or objects. Except in
this sense of the points of the P-sets or objects of the space
${\rm\bf B}$, it is not clear as to whether we can, in any other
alternative manner, ``mathematically'' recover the non-singular,
physical, notion of a particle to represent material bodies. In
the absence of any other alternatives, these concepts related to
dynamical systems on the space ${\rm\bf B}$ then become logical
compulsion.

In conclusion, from the considerations of the present paper,
conceptions related to dynamical systems on the space ${\rm\bf B}$
appear logically inevitable. The unification of the conceptions of
the quantum theory and those of the general theory of relativity
as embodied in the formalism of the dynamical systems on the space
${\rm\bf B}$, then, becomes {\em natural\/} from this point of
view.


\begin{thebibliography}{99}
\bibitem{heisenberg} Heisenberg W (1949) {\it The Physical
principles of the quantum theory\/} (Dover, New York)

\bibitem{cartan} A classic reference is: Cartan E (1923)
{\it Ann. Ec. Norm.}, {\bf 40}, 325; (1924) {\bf 41}, 1

\bibitem{class-mech} See, for example, Goldstein H (1950) {\it Classical Mechanics\/} (John
Wiley \& Sons, New York) \\ Kibble T W B (1970) {\it
Classical Mechanics\/} (ELBS-McGraw-Hill, London) \\
Sudarshan E C G and Mukunda N (1974) {\it Classical Dynamics - A
modern perspective}, (Wiley Interscience, New York)

\bibitem{bohr1} Bohr N (1928) {\it Nature (Suppliment Series)}, April 14,
1928, p. 580

\bibitem{ein-pop}  Einstein A (1968) {\it Relativity: The Special
and the General Theory\/} (Methuen \& Co. Ltd, London) (See, in
particular, Appendix V: Relativity and the Problem of Space.)

\bibitem{varadrajan} Varadrajan V S (1988) {\it Geometry of
quantum theory}, Vol. I and II (Van Nostrand Reinhold Company, New
York)

\bibitem{bohr2} Bohr N (1970) in {\it Albert Einstein: Philosopher
Scientist} (Ed. P A Schlipp, La Salle: Open Court Publishing
Company - The Library of Living Philosophers, Vol. VII: 1970)

\bibitem{jackson} Jackson J D (1968) {\it Classical
Electrodynamics} (Academic Press)

\bibitem{dirac} Dirac P A M (1970) {\it Principles of Quantum
Theory\/} (Dover, New York) \\ Scheweber J (1980) {\it
Relativistic Quantum Theory\/} (McGraw Hill, New York)

\bibitem{subtle} Pais A (1982) {\it Subtle is the Lord ... The
science and the life of Albert Einstein} (Oxford: Clarendon
Press)

\bibitem{std-texts} See, for example, Weinberg S (1972) {\it
Gravitatioan and Cosmology} (Wiley, New York) \\ Misner C, Thorne
K S and Wheeler J A (1973) {\it Gravitation} (Freeman, San
Francisco) \\ Davies P C W (1986) {\it Quantum Fields in Curved
Spacetime\/} (Cambridge University Press, Cambridge)

\bibitem{smw-field} Wagh S M (2004) {\it Einsteinian field theory as a program
in fundamental physics} {\bf Database: physics/0404028} and
references therein

\bibitem{schlipp} Einstein A (1970) in {\it Albert Einstein:
Philosopher Scientist} (Ed. P A Schlipp, La Salle: Open Court
Publishing Company - The Library of Living Philosophers, Vol VII).

\bibitem{smw-issues} Wagh S M (2004) {\it Some fundamental issues in
General Relativity and their resolution} {\bf Database:
gr-qc/0402003} and references therein

\bibitem{kdjoshi} Joshi K D (1983) {\it Introduction to General
Topology} (Wiley Eastern, New Delhi)\\
Lipschutz S (1981) {\em Theory \& Problems of General Topology}
(Schaum's Outline Series, McGraw-Hill International, Singapore)

\bibitem{trim6} Nadkarni M G (1995) {\it Basic Ergodic Theory}
(Texts and Readings in Mathematics - 6: Hindustan Book Agency, New
Delhi) and references therein

\bibitem{smw-indeterminacy} Wagh S M (2004) {\it On the continuum
origin of Heisenberg's indeterminacy relations} {\bf Database:
physics/0404066}. Submitted to GRG.

\end{thebibliography}
\end{document}